\documentclass[10pt]{article}
\usepackage{amsfonts}
\begin{document}

\def\wh{\widehat}
\def\wt{\widetilde}
\def\D{{D}}
\def\np{\not{{P}}}
\def\ov{\overline}
\def\udt#1{$\underline{\smash{\hbox{#1}}}$}
\def\noi{\noindent}

\centerline{\bf Gauge invariant formulation of systems with second class
constraints.}
\vskip .2cm
\centerline{{\udt{J.Stephany} } and  A. Restuccia}
\vskip .2cm
\centerline{\it Universidad Sim\'{o}n Bol\'{\i}var, Departamento de F\'{\i}sica}
\centerline{\it Apartado postal 89000, Caracas 1080-A, Venezuela.}
\vskip .3cm

The covariant quantization of physical systems with reducible constraints is
one of the unresolved problems which appear in connection with the Green-Schwarz
Superstring (GSSS).The formulation of the GSSS [1] presents several  advantages
over the perturbative NRS formulation [2]. In particular a non  perturbative
second quantized Light Cone Gauge formulation was obtained in [1]and the closure
of the Super-Poincare algebra [3] and multiloop analysis  of the S-matrix was
performed in [4]. Nevertheless the  covariant first quantization of the GSSS
is a problem which has not been solved. The main difficulty has been  the
covariant gauge fixing of the local kappa-Supersymmetry [5]. In fact, the  first
class constraints associated to the gauge symmetries appear mixed with  second
class ones, and so far no local, Lorentz covariant and finite reducible  [6] [7]
approach to disentangle them has been found.The zero mode structure of GSSS is
described by the Brink-Schwarz Superparticle, (BSSP) [8]. The canonical structure
of both  theories presents a very close correspondence. In particular first and
second  class constraints in the BSSP also appear mixed and similar to the
GSSS, problems  with the infinite reducibility of the covariant generators of
gauge  symmetries are present.

Here we present a canonical approach which applied to the covariant
quantization of the BSSP and resolves the above mentioned problems. The
formulation is based upon a general canonical approach  for dynamical systems
restricted by reducible first and second class  constraints [9]. In this
approach, which is closely related to the work in [10],  the phase space is
extended to a larger manifold where all extended constraints  are first class.
By an appropriate gauge fixing one may reduce the functional  integral to a
functional integral on the original constrained manifold, with  the correct
functional measure. We show explicitly the reduction procedure. It  is an
off-shell approach allowing the systematic construction of the off-shell
nilpotent BRST charge and of the BRST invariant effective action. For the BSSP
 this approach leads directly to the correct BRST charge. It gives a systematic
method for the construction of the superparticle action, proposed in Ref.[11] by
Kallosh which is related to this charge through the BFV formalism [12].

We start with  constrained system with  Hamiltonian $H_0$
subject to a set of reducible constraints $\phi_{a_1}$  $(a_1 =1,\cdots ,n)$ and
a set of first class constraints  $\varphi_i$ $(i=1,\cdots ,k)$ which we omit in
the explicit construction that  follows. We limit ourselves to remark on the
modifications to be done when  included. So we have,
\begin{equation}
\label{1}
\phi_{a_1}=0
\end{equation}
\begin{equation}
\label{2}
a_{a_2}^{a_1}\phi_{a_1}=0\ \ \ \ \ \ \ a_1=1\cdots n,\ \ \ a_2=1\cdots m
\ \ \ \ \ .
\end{equation}

We will not suppose $a_{a_2}{}^{a_1}$ to be of maximal rank. Instead we will
impose that a (T+L) decomposition  is allowed.

We have then for any objects $V_{a_1}$ and $W^{a_1}$
\begin{eqnarray}
\label{3}
V_{a_1} =V^\top_{a_1}+A_{a_1}^{a_2}V_{a_2}^L\ \nonumber \\
a^{a_1}_{a_2}V_{a_1}^\top =0,\ \ \ \ \ \ \ V_{a_2}^L=a^{a_1}_{a_2}V_{a_1} \\
W^{a_1} =W_\top^{a_1}+a^{a_1}_{a_2}W_L^{a_2} \\
A^{a_2}_{a_1}W_\top^{a_1}=0,\ \ \ \ \ \ \ W_L^{a_2}=A^{a_2}_{a_1}W_{a_1}\ \ \ \ \ . \nonumber
\end{eqnarray}

It follows that $V_{a_2}^L = a^{a_1}_{a_2}A^{b_2}_{a_1}V_{b_2}^{L}$,
$W_L^{a_2}=A^{a_2}_{a_1}a^{a_1}_{b_2}W_L^{b_2}$ and $W^{a_1}V_{a_1} =W^{a_1}_\top
V_{a_1}^\top +W_L^{a_2} V^L_{a_2}$

In the irreducible case  $A^{a_2}_{a_1}$ is the inverse of $a_{a_2}^{a_1}$. In
the finite reducible case this  decomposition may always be done in a unique way
for a given pair $A,a$.  For infinite reducible
system, we will assume that there exists such a decomposition.The constraints
(2) are second class in the sense that they have an invertible  Poisson Bracket
matrix in the transverse sub-space.

Following Ref. 9 and 10 let us enlarge the phase space using a set of
auxiliary variables $\xi^{a_1}$ and $\eta_{b_1}$ conjugate to each other.
We also introduce the combinations
\begin{equation}
\label{4}
\Phi_{a_1} =\eta_{a_1}-{1\over{2}}\omega_{a_1b_1}(p,q)\xi^{b_1}\\
\ov{\Phi}_{a_1} =\eta_{a_1}+{1\over{2}}\omega_{a_1b_1}(p,q)\xi^{b_1}
\ \ \ \ \ .
\end{equation}

Here $\omega_{ab}$ is an antisymmetric matrix with vanishing Poisson Bracket with
itself to be fixed by the procedure. $\ov{\Phi}$ and $\Phi$ satisfy
\begin{equation}
\label{5}
\{\Phi_{a_1},\Phi_{b_1}\}\ =-\omega_{a_1b_1}\\
\{\ov{\Phi}_{a_1},\ov{\Phi}_{b_1}\}\ =\omega_{a_1b_1}\\
\{\Phi_{a_1},\ov{\Phi}_{b_1}\}\ =0\ \ \ \ \ .
\end{equation}
In order to introduce only the complications necessary in  the case of
the BS superparticle we will suppose in the following that $\omega_{a_1b_1}$ is
transverse
\begin{equation}
\label{6}
a^{a_1}_{a_2}\omega_{a_1b_1}=0\ \ \ \ \ \ a_1=1\cdots n
\ \ \ \ \ \ a_2=1\cdots m
\end{equation}
and invertible in the transverse space.

Now let us extend the constraints in the enlarged space to
\begin{equation}
\label{7}
\wt{\phi}_{a_1}=\phi_{a_1}+V^{c_1}_{a_1}\Phi_{c_1}=0
\end{equation}
where  $V^{c_1}_{a_1}(q,p)$ is also to be fixed. In general the first class
constraints $\varphi$ may  also have to be extended in order the complete set
of extended constrains be first class. In this case of the superparticle,
however, the extension is not necessary. We assume $V_{a_1}^{b_1}$ to
be invertible. In this case we impose the constraints (7) to be irreducible,
first class and with structure functions at most linear in $\Phi_{a_1}$. We
then have
\begin{equation}
\label{8}
\{\wt{\phi}_{a_1},\wt{\phi}_{b_1}\}=U_{a_1b_1}^{c_1}\wt{\phi}_{c_1}
=-2(u_{a_1b_1}^{c_1}+v_{a_1b_1}^{c_1d_1} \Phi_{d_1})\wt{\phi}_{c_1} .
\end{equation}

The structure functions $U_{a_1b_1}^{c_1}$ may depend on the phase space
variables $p$ and $q$. Substitution of (7) in (8) yields.
\begin{eqnarray}
\label{9}
\{\phi_{a_1},\phi_{b_1}\} -V_{a_1}^{c_1}V_{b_1}^{d_1}\omega_{c_1d_1}+
2u_{a_1b_1}^{c_1}\phi_{c_1}=0  \nonumber \\
\{\phi_{a_1},V_{b_1}^{c_1}\}+\{V_{a_1}^{c_1},\phi_{b_1},\}
+2v_{a_1b_1}^{d_1c_1}\phi_{d_1}+ 2u_{a_1b_1}^{d_1}V_{d_1}^{c_1}=0 \\
\{V_{a_1}^{c_1},V_{b_1}^{d_1}\}+\{V_{a_1}^{d_1},V_{b_1}^{c_1}\}
+2V_{e_1}^{c_1}v_{a_1b_1}^{e_1d_1}+ 2V_{e_1}^{d_1}v_{a_1b_1}^{e_1c_1}=0\ \ \ \ . \nonumber
\end{eqnarray}

We suppose here that
\begin{equation}
\label{10}
\{\phi_{a_1},\Phi_{1a_1}\}=0,\ \ \ \{V_{1a_1}^{b_1},\Phi_{1c_1}\}=0
\ \ \ \ \ .
\end{equation}

Let us suppose that we are able to find a solution to (20) with all the
required conditions. In order to demonstrate the equivalence of our system in
the enlarged phase  space to the original system we have to impose additional
restriction besides  (7). A counting of the degrees of freedom suggests which
ones should be  chosen in this generalized situation. The original model has
$2N$ phase  space variables $p,q$ restricted by ($n-m_L$) transverse constraints
with  $m_L$ the rank of $a^{a_2}_{a_1}$. The enlarged model has $2N$ variables
$p,q$ and $2n$ variables $\xi$, $\eta$ restricted by $n$ constraints
$\wt{\phi}_{a_1}$ and $n$ gauge fixing conditions $\wt{\chi}_{a_1}$. To match
we need ($n-m_L$) additional constraints. We take them to be
\begin{equation}
\label{11}
\ov{\Phi}_{a_1}^\top =0\ \ \ \ \ .
\end{equation}
Since
\begin{equation}
\label{12}
[\ov{\Phi}_{a_1}^\top ,\ov{\Phi}_{a_2}^\top ]=\omega_{a_1a_2}^\top
\end{equation}
the constraints (11) are in our hypothesis second class.The advantage of
this formulation is that he field dependence in $\omega^{\top}_{a_1a_2}$   may
be simpler than in $\{\phi_{a_1},\phi_{b_1}\}$ since  $V_{a_1}^{a_2}$ may be
also a field dependent object.

A gauge invariant extension of the hamiltonian $H_0$ may be written in the
form [9]
\begin{equation}
\label{13}
\wt{H}=H_0+h^{a_1}\Phi_{a_1}\ \ \ \ \ .
\end{equation}
$h^{a_1}$ is fixed imposing
\begin{equation}
\label{14}
\{\wt{H},\wt{\phi}_{a_1}\} =W_{a_1}^{b_1}\wt{\phi}_{b_1}\ \ \ \ \ .
\end{equation}

Introducing the ghost variables $C^{a_1}$ and $\mu_{a_1}$ the BRST operator
and the extended hamiltonian are obtained in the standard way [6]

The BRST invariant effective action in a phase space representation is given
by [6] [18]
\begin{equation}
\label{15}
S_{eff}=<p\dot{q}+\mu_{a_1}\dot{C}^{a_1}+\eta_{a_1}\dot{\xi}^{a_1}-
\wh{H}+\wh{\delta}(\lambda^{a_1}\mu_{a_1})+\wh{\delta}(\ov{C}_{a_1}\chi^{a_1})>
\end{equation}
where $\chi^{a_1}$ are the gauge fixing conditions and $\wh{\delta}$ is
defined by $\wh{\delta}F=[\Omega ,F]$
for any function $F$ of the canonical variables of
the enlarged superphase-space. For the non-canonical sector we have
\begin{equation}
\label{16}
\wh{\delta}\lambda^{a_1} =\theta^{a_1}\ \ \ ,\ \wh{\delta}\theta^{a_1}=0
\\  \wh{\delta}\ov{C}_{a_1} =B_{a_1}\ \ \ ,\ \wh{\delta}B_{a_1}=0 \\
\delta\mu_{a_1}=\wt{\phi}_{a_1}\ \ \ \ \ .
\end{equation}

We will show  now that with an adequate gauge fixing condition one can reduce
the path  integral corresponding to the enlarged system to the
Senjanovic-Fradkin  expression for the original system.

We may  choose the gauge conditions
\begin{equation}
\label{17}
\chi^{a_1}=\xi^{a_1}\ \ \ \ \ .
\end{equation}

The functional integral is
\begin{equation}
\label{18}
I(\chi )=\int \D z\delta (\ov{\Phi}^\top )(det \omega^\top
)^{1/2}e^{-Seff}
\end{equation}
where $\D z$ is the Liouville measure

Integrating in $\theta$ one gets $\delta (\mu )$.Integrating in $B_{a_1}$,
$\ov{C}_{a_1}$,and $\lambda^{a_2}_L$ and using Eq.(9) the factor in the
measure of (18) becomes

\begin{equation}
\label{20}
(det \omega^\top )^{1/2}\delta (\eta )\delta (\xi )\delta (C)det V_{\top}^\top
\ \ \ \ \ .
\end{equation}

Now we note from (9) that
\begin{equation}
\label{21}
(det \omega^\top )^{1/2}det V_{\top}^\top =(det \{\phi_\top ,\phi_\top \} )
\ \ \ \ \ .
\end{equation}

Doing the trivial integrals in $\eta$, $\xi$ and $C$, we  finally obtain
\begin{equation}
\label{22}
I=\int \D q \D p \D \lambda^\top (det \{\phi_\top ,\phi_\top \})exp-
<p\dot{q}-H+\lambda_\top \phi^\top >
\end{equation}
which is the correct Senjanovic-Fradkin expression of the functional integral
of this system.

\end{document}